\documentclass[]{interact}

\usepackage{epstopdf}% To incorporate .eps illustrations using PDFLaTeX, etc.
\usepackage[caption=false]{subfig}% Support for small, `sub' figures and tables
\usepackage{subfiles}
\usepackage{array}
\usepackage{enumitem}
\usepackage{makecell}
\usepackage{colortbl}
\usepackage[table]{xcolor}
\definecolor{siggreen}{RGB}{198, 239, 206}  % light green
\usepackage[T1]{fontenc}
\usepackage{tikz}
\usetikzlibrary{
    positioning,
    backgrounds,
    fit,
    arrows.meta,
    patterns,
    calc,
    shapes.geometric
    }

\usepackage[natbibapa,nodoi]{apacite}% Citation support using apacite.sty. Commands using natbib.sty MUST be deactivated first!
\begin{document}

\articletype{RESEARCH ARTICLE}% Specify the article type or omit as appropriate

\title{Agentic AI Autonomy Assessment: A Decision-Support Framework Towards Governed Supply Chain Systems}

\author{
\name{Lennart Trumpler\textsuperscript{a}\thanks{Corresponding author: Lennart Trumpler. e-mail: petru@iti.sdu.dk}, Rodrigo Furlan de Assis\textsuperscript{a}, Elias Ribeiro da Silva\textsuperscript{a}, Luis Antonio de Santa-Eulalia\textsuperscript{b}, Christian Hendriksen\textsuperscript{c}}
\affil{\textsuperscript{a}Center for Supply Chain Digitalisation, University of Southern Denmark, Sønderborg, Denmark; \\
\textsuperscript{b}École de Gestión, Université de Sherbrooke, Sherbrooke, Canada; \\
\textsuperscript{c}Department of Operations Management, Copenhagen Business School, Copenhagen, Denmark}
}

\maketitle

\begin{abstract}
Supply chain decision-making is rapidly transforming with the rise of agentic AI – highly autonomous systems that can operate on complex, long-horizon tasks. Yet the adoption of agentic systems outpaces their governance: existing taxonomies of autonomy only offer discrete classifications, rely on subjective judgement, and cannot track autonomy across a system's life cycle, leaving enterprises unable to assess the risks of increasingly autonomous supply chain agents. This paper proposes the Agentic AI Autonomy Assessment (AAAA) framework, which defines and measures the degree of autonomy at a task level. The framework is based on the three dimensions of user delegation, consultation, and collaboration, enabling continuous monitoring from an agent's development through its runtime to end-of-life. The framework's construct validity was tested in a simulated beer distribution game, examining how the autonomy score relates to a company's performance. Results reveal a weak link between autonomy and tier costs with a positional effect: upstream tiers benefit from higher autonomy while downstream tiers are harmed, positioning autonomy as an inherent dimension of agentic systems, orthogonal to capability. The framework provides a foundation for risk assessment, governance, and transparent autonomy policies to support the governed enterprise adoption of agentic AI in supply chains.
\end{abstract}

\begin{keywords}
Autonomy measurement; supply chain governance; multi-agent systems; large language models; human-AI collaboration
\end{keywords}

%----- MAIN BODY

\section{Introduction}

% -----  INTRO
As supply chains form the backbone of virtually every industry, there is a growing need for them to become more flexible and resilient \citep{xu_multi-agent_2024}. In this context, the fast-paced digitalisation of global supply chains is compelling decision-makers to continuously transform their value chains by adopting new digital capabilities to sustain a competitive advantage. At the same time, supply chains have become increasingly complex over the last few decades, creating significant challenges for supply chain managers. Although digital technologies offer promising opportunities to address these challenges, their full potential remains untapped in practice \citep{sengupta_impact_2025}. 

% -----  MOTIVATION - AGENTS, MAS
This complexity shift has intensified the need for computational approaches capable of supporting decentralised decision-making under uncertainty. For decades, the dominant response to this problem has been traditional analytical models, but they often struggle to capture the distributed nature of supply chain actors, their interdependencies, and the dynamic interactions that emerge across multiple echelons \citep{choi_supply_2001}. In response, agents, agent-based systems, and in particular multi-agent systems (MAS) emerged and have been extensively investigated over the past decades \citep{swaminathan_modeling_1998, xu_multi-agent_2024, liu_multi-agent_2025}. In this paradigm, agents represent autonomous or semi-autonomous, distributed decision-makers that are goal-driven and interact with one another to achieve system-level objectives \citep{wooldridge_intelligent_1995}. As a result, MAS has become a consolidated approach for modelling autonomous supply chain entities and addressing the inherent complexity of supply chain systems. By distributing decision-making and computational efforts among multiple agents, MAS could therefore handle larger and more complex problems than traditional centralised approaches. However, due to the prior limitations within AI systems, these systems often acted under strict rules and constraints, demonstrated only rudimentary intelligence, exhibited limited autonomy, minimal adaptability to their environment \citep{xu2024, sapkota2025}.

% -----  AGENTIC AI
Following the development of generative AI and Large Language Models (LLM), recent technological advancements have introduced the term 'agentic AI', a new paradigm highlighting the rapid advancement in AI capabilities and the resulting surge in highly autonomous systems that can operate independently on complex tasks over long periods of time \citep{kwa_measuring_2026, sapkota2025}. This is now possible given that capabilities, including reasoning and agency, are more easily embedded into the agents, which constitute one of the recent technological leaps expected to impact large parts of society and industries \citep{acharya_agentic_2025}. 

% -----  INDUSTRY APPLICATIONS
Current industry applications are being rapidly pushed to market, including companies such as PactumAI, that already offer agentic solutions to major companies like Walmart by automating a large tail of supplier negotiations through agentic systems. With AI assistants like OpenAI's ChatGPT \citep{openai2025chatgpt} or Anthropic's Claude \citep{anthropic_claude_2026} increasingly adopting agentic capabilities, these systems quickly move to become more accessible and consequently active in supply chain decision-making, thus shaping supply chain dynamics \citep{hendriksen_artificial_2023}.

% -----  GAP
However, the rapid adoption of agentic AI systems leaves behind a gap in the current body of research on how these systems can be adopted without compromising governance \citep{dwivedi_opinion_2023}. This requires the development and integration of new constructs for AI autonomy and the development of new benchmarks for the supply chain context \citep{hendriksen_artificial_2023, dubey_benchmarking_2024}. 

While the literature has offered and discussed taxonomy frameworks dealing with levels of autonomy in AI and supply chains \citep{feng_levels_2025, mitchell_fully_2025, xu2024}, the existing approaches fall short when applied to this new agentic context. First, the taxonomies only offer discrete autonomy classes, hindering a more nuanced assessment of autonomy. Second, the assessments only provide heuristic guidance, which leaves room for interpretation and risks of introducing human biases in the judgment. Finally, existing literature often argues that autonomy is an emergent behaviour rather than a characteristic that can already be accounted for in the development stage \citep{bandi_rise_2025}. 

% -----  RESEARCH QUESTION AND APPROACH
Hence, the rising adoption of agentic AI in the supply chain context, paired with the theoretical gaps in supply chain literature on proper autonomy levels of agents, creates major risks for disruptions \citep{hendriksen_artificial_2023} while unable to demonstrate how performance gains can be sustained over time. Specifically, the literature has yet to address how autonomy in agentic AI-based supply chain systems can be systematically defined, allocated, assessed, and monitored over time across diverse supply chain stakeholders to ensure effective system governance and governability. This study, therefore, focuses on the following open research question: \\

% -----  RESEARCH QUESTION
\textit{RQ: How can the degree of autonomy of agentic AI systems be defined, measured objectively, and monitored continuously to support their governed adoption in supply chains?} \\

% -----  CONTRIBUTIONS
By answering this question, the study contributes to the body of knowledge in supply chain theories on AI adoption by creating a framework for agentic AI autonomy assessment. The framework can be applied throughout the full agentic AI system life cycle, ranging from the development stage, runtime, to the end-of-life. By offering a way to continuously monitor the system's autonomy, the governance surrounding the integration of agentic AI into supply chains can be improved, addressing the risks of disruption that highly autonomous systems currently pose \citep{hendriksen_artificial_2023}, such as loss of human control, goal misalignment, unpredictable emergent behavior, coordination risks, and so forth. Practically, the framework enables the development of more transparent AI policies, such as guardrails or boundaries for the system's autonomy that integrate with research on human-AI collaboration. 

% -----  ROADMAP
The remaining study is divided into the following sections: Section \ref{sec:background} provides a comprehensive literature review on the topic of agentic AI in supply chains and how autonomy relates to human-AI decision making. Section \ref{sec:aaaa_framework} presents the novel Agentic AI Autonomy Assessment (AAAA) framework, which enables the measurement of an autonomy score. Section \ref{sec:validation} applies the framework to the context of a simulated instance of the beer distribution game, presenting how the autonomy score relates to system performance in cost and fill rate. \ref{sec:Dicussion} discusses the current construct validity and results of the experiments. Section \ref{sec:Conclusion} presents the conclusion and dives deeper into the theoretical contributions and practical implications of our findings.

\section{Background}
\label{sec:background}

\subsection{Agentic AI in Supply Chains and Operations Management}
Supply chains have long been recognised as complex distributed systems composed of multiple autonomous decision-making entities operating under uncertainty, incomplete information, and conflicting objectives \citep{ivanov2019}. Decisions regarding procurement, production, transportation, inventory, and distribution are often made locally while simultaneously influencing the performance of the entire network \citep{bottani2019b}. This distributed nature has motivated decades of research on autonomous agents, agent-based systems, and multi-agent systems (MAS) as computational paradigms for representing decentralized decision-makers capable of coordinating planning, negotiation, scheduling, and resource allocation across supply chain networks \citep{xu2021}.

Traditional MAS demonstrated significant potential for modelling decentralised supply chain decision-making. However, their practical adoption remained constrained by several limitations \citep{camarillo2018}. Most architectures relied on predefined communication protocols, handcrafted decision rules, structured data representations, and explicitly programmed coordination mechanisms \citep{xu2024a}. As a consequence, these systems often struggled to operate effectively in dynamic environments requiring flexible reasoning, adaptation to unforeseen situations, and the interpretation of heterogeneous or unstructured information \citep{hughes2025, xu2024}.

Recent advances in foundation models and generative AI have substantially expanded the capabilities of autonomous agents \citep{alva2026}. By integrating LLM with memory, planning, reasoning, tool utilization, and workflow orchestration, agents are increasingly capable of understanding natural language, interacting with external systems, decomposing complex objectives into executable tasks, and adapting their behaviour to changing operational conditions \citep{trifone2026}. Rather than replacing the principles established by traditional MAS, these developments extend them, enabling more flexible and cognitively capable agents that can operate in open and uncertain environments \citep{islam2026}.

This evolution has led to the emergence of Agentic AI, where autonomous agents no longer operate as isolated software components but as goal-driven entities capable of reasoning, coordinating with other agents, interacting with humans, and executing multi-step workflows across dynamic environments \citep{hughes2025, ren2025}. Unlike conventional AI applications that primarily generate predictions or recommendations for human interpretation, agentic systems increasingly participate directly in operational and tactical decision-making processes \citep{sapkota2025}. Consequently, AI is progressively evolving from an analytical support technology toward an active decision-making entity within supply chain operations \citep{parthasarathy2026}.

\subsection{Decision Authority and Delegation}

The growing adoption of agentic AI in supply chains is accompanied by a progressive transfer of decision authority from human actors to artificial agents \citep{acharya_agentic_2025}. Unlike traditional decision-support systems, which primarily provide recommendations, agentic systems can interpret objectives, initiate actions, coordinate resources, and execute operational tasks with varying levels of independence \citep{hughes2025}.

This transition has shifted the discussion from automation capabilities toward authority allocation. In this context, autonomy can be understood as the degree to which an agent can operate without the involvement of another entity, whether human or artificial. However, delegated decision authority should not be equated with autonomy itself \citep{fuchs2023}. Authority determines which decisions an agent is permitted to make, whereas autonomy reflects how independently those decisions are actually executed in practice \citep{morris2023,feng_levels_2025}. Consequently, organizations must establish governance mechanisms defining when authority can be delegated, when human approval is required, and how responsibility is distributed across decision processes \citep{marquez2026}.

Rather than representing a binary distinction between human and machine control, recent research increasingly conceptualises decision authority as a continuum in which responsibility can be distributed across humans and artificial agents according to the characteristics of the decision context \citep{feng_levels_2025}. Under this perspective, authority allocation becomes a design choice rather than a fixed system property, allowing different levels of human involvement throughout the decision process \citep{zhang2026decision}. Emerging concepts such as negotiated agency and adjustable autonomy further argue that the appropriate allocation of authority should consider factors such as task complexity, operational risk, and governance requirements \citep{leonardi2025}.

The literature further suggests that authority can be delegated across different stages of organisational decision-making rather than being limited to task execution alone \citep{hasselwander2026}. Table~\ref{tab:delegation_forms} summarises representative forms of delegated authority identified in recent agentic AI research.

\begin{table}[ht]
\tbl{Representative forms of authority allocation in agentic AI systems}{
\begin{tabular}{p{1.8cm} p{3.1cm} p{3.1cm} p{2.1cm} p{2.1cm}}
\toprule
\textbf{Decision Stage} &
\textbf{Delegated Authority} &
\textbf{Representative Supply Chain Activities} &
\textbf{Typical Human Role} &
\textbf{Representative Sources} \\
\toprule

Situation Assessment &
Interpret, prioritise and contextualise operational information &
Demand anomaly detection, disruption monitoring, supplier risk identification &
Validate interpretations and define objectives &
\cite{acharya_agentic_2025}\\

\midrule

Planning and Execution &
Generate plans and execute operational decisions within predefined boundaries &
Inventory replenishment, production scheduling, transportation planning &
Define constraints and supervise execution &
\cite{el2026}\\

\midrule

Coordination &
Allocate tasks, synchronise activities and coordinate multiple decision-makers &
Multi-agent orchestration, exception management, collaborative planning &
Monitor objectives and intervene when necessary &
\cite{jannelli_agentic_2025}\\

\bottomrule
\end{tabular}}
\label{tab:delegation_forms}
\end{table}

Table~\ref{tab:delegation_forms} illustrates that authority allocation extends beyond operational execution. During situation assessment, authority is delegated through the interpretation and prioritization of information \citep{acharya_agentic_2025}. During planning and execution, agents influence operational outcomes by generating plans and selecting actions within predefined constraints \citep{el2026}. Finally, coordination-level delegation grants authority over the distribution and synchronization of activities across multiple actors, representing the highest degree of influence over system behaviour \citep{jannelli_agentic_2025}. 

Despite increasing levels of delegated authority, recent studies recognise that the allocation of decision authority introduces important governance challenges, particularly regarding accountability, oversight, and responsibility \citep{narayana2025}. Rather than eliminating human involvement, agentic systems are frequently designed to support different forms of human oversight, including approval mechanisms, intervention points, and monitoring activities \citep{abou2025}. Consequently, the allocation of authority must be considered together with the governance mechanisms that regulate when and how human actors remain involved in decision-making processes \citep{murugesan2025}.

The allocation of decision authority also depends on the extent to which agent behaviour can be understood and supervised \citep{ciardo2020attribution}. Transparency and explainability are therefore frequently discussed as enabling mechanisms that support human oversight and facilitate the inspection, justification, and, when necessary, revision of agent decisions \citep{hosseini2025}. Because decision authority may be distributed across multiple stages of the decision process, governance is not limited to deciding whether an agent is autonomous, but also to determining where human oversight should be maintained throughout goal execution \citep{leonardi2025}.

\subsection{Goal Decomposition and Task Formation}

Understanding autonomy in agentic systems requires understanding how goals are interpreted, decomposed, and transformed into executable tasks \citep{kumar2026, jannelli_agentic_2025}. While traditional automation systems rely on predefined workflows designed during system development, agentic systems possess the ability to dynamically generate, modify, and allocate tasks during execution, enabling them to pursue complex objectives in uncertain and evolving environments \citep{abou2025}.

Goal decomposition refers to the process of transforming a high-level objective into a hierarchy of manageable sub-goals and tasks \citep{khot2022}. This capability is frequently identified as one of the defining characteristics of agentic systems, distinguishing them from traditional AI applications that primarily respond to isolated requests. Rather than solving a single well-defined problem, agentic systems can pursue long-term horizon objectives that require planning, intermediate reasoning steps, and the coordination of multiple activities over time. Recent studies describe this process as a form of hierarchical planning in which a global objective is progressively refined into operational activities that can be executed by specialized agents or external tools \citep{prasad2024, bidochko2025}.

In multi-agent architectures, goal decomposition is often coordinated by an orchestrator agent responsible for receiving high-level objectives and distributing sub-goals across specialised agents \citep{fernandes2025}. Although the underlying technical implementations may differ, the common characteristic across these approaches is the ability to create structured plans that connect strategic objectives with operational actions \citep{he2026}.

Closely related to goal decomposition is task formation, which refers to the generation of executable actions required to achieve a given sub-goal \citep{majumder2023}. While goal decomposition determines what should be accomplished, task formation determines how objectives are operationalized \citep{nowaczyk2025}. Recent agentic architectures support adaptive task formation, enabling agents to generate, modify, reorder, and delegate tasks in response to changing environmental conditions \citep{adam2024, abou2025}. Besides, unlike traditional business process automation systems, where workflows are predefined, agentic systems can dynamically construct and revise execution plans during runtime, allowing them to operate under uncertainty and evolving operational conditions \citep{yao_react_2023}.

The extent to which agents participate in goal decomposition and task formation depends on the authority allocated to the system \citep{lee2026}. In low-autonomy configurations, humans retain responsibility for planning, task design, and task allocation, while agents primarily execute predefined activities. As autonomy increases, agents progressively assume responsibility for generating plans, creating tasks, and coordinating execution \citep{feng_levels_2025}. Consequently, autonomy is expressed not only through the execution of tasks, but also through the ability to formulate, structure, and adapt the tasks required to achieve a given objective \citep{kasirzadeh2025}. 

\begin{figure}[ht]
\centering
\includegraphics[width=\linewidth]{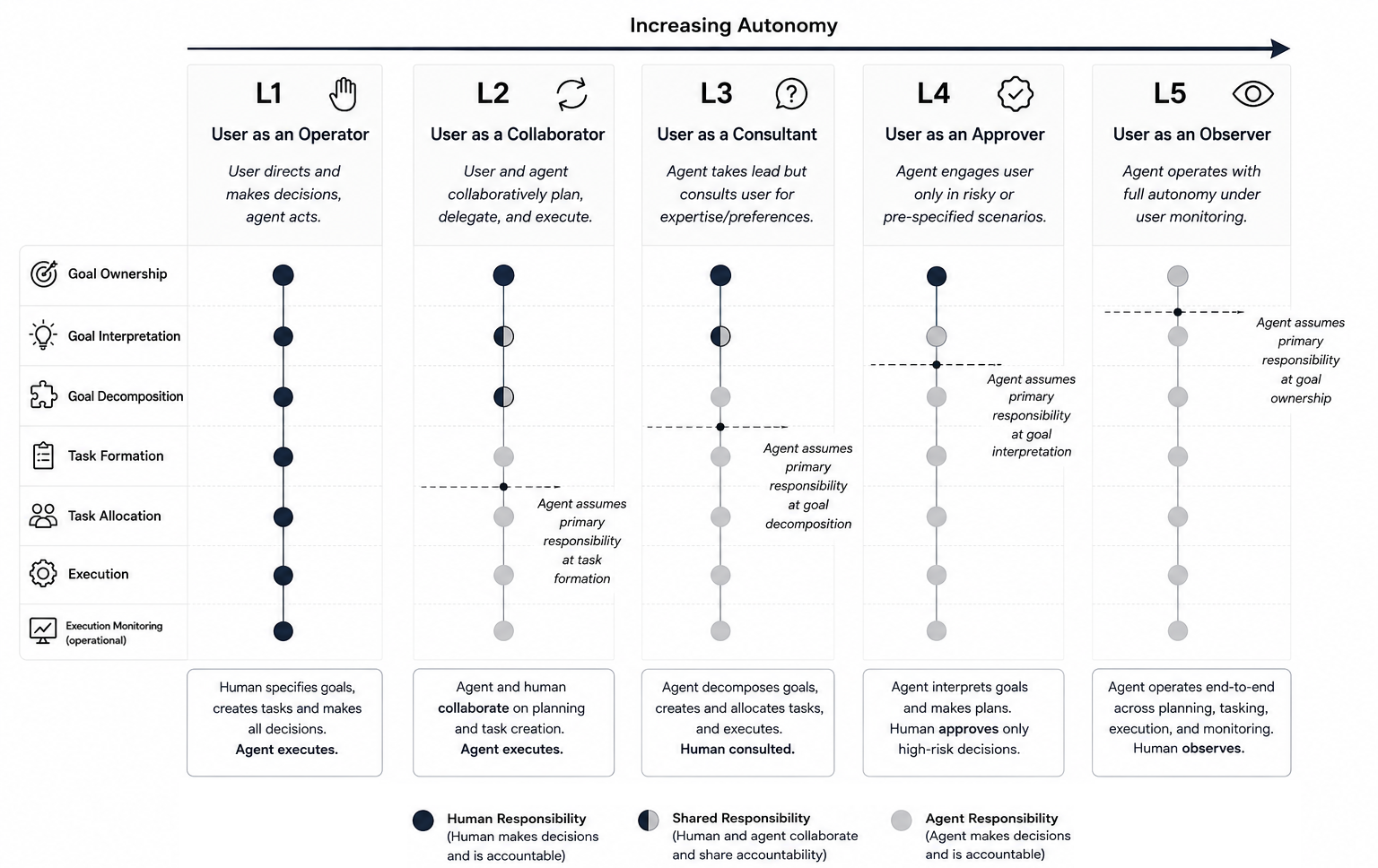}
\caption{Distribution of human, shared, and agent responsibility across stages of the goal realization process under different autonomy configurations. The horizontal marker indicates the stage at which the agent assumes primary decision responsibility. Adapted from autonomy-level frameworks proposed in \cite{morris2023,feng_levels_2025,kasirzadeh2025}.}
\label{fig:goal_decomposition_autonomy}
\end{figure}

Figure \ref{fig:goal_decomposition_autonomy} illustrates this progression by showing how responsibility for each stage of the goal realization process shifts between human and artificial actors across different autonomy configurations. The figure also identifies the stage at which the agent assumes primary responsibility for decision-making in each autonomy level. Adapted from existing autonomy-level frameworks \cite{morris2023,feng_levels_2025,kasirzadeh2025}, the figure emphasizes that autonomy extends beyond task execution to include the degree to which agents participate in planning and task generation.

Figure \ref{fig:goal_decomposition_autonomy} illustrates how increasing autonomy progressively transfers responsibility from humans to agents throughout the goal realization process. While responsibility may be shared during intermediate planning activities, the horizontal marker in each autonomy configuration identifies the stage at which the agent assumes primary decision responsibility. Goal ownership remains with the human across all autonomy levels; however, primary responsibility for transforming high-level objectives into executable actions progressively shifts from execution (L1), to task formation (L2), goal decomposition (L3), and goal interpretation (L4–L5). Consequently, increasing autonomy is characterized not only by greater independence during task execution, but also by an expanding ability of agents to interpret objectives, generate intermediate goals, formulate executable tasks, coordinate their execution, and monitor operational outcomes with progressively less human involvement. Autonomy is therefore understood not only in terms of who executes actions, but also in terms of who generates, structures, and coordinates the work to be executed, making the provenance of goals and tasks a central dimension for assessing autonomy in agentic systems.

\subsection{Distributed Decision Execution in Agentic Architectures}

The previous sections established that agentic AI changes how decision authority is allocated between humans and artificial agents and how this authority becomes operational through goal decomposition and task formation. However, the execution of decomposed goals rarely occurs within a single decision-making entity \citep{bidochko2025}. In supply chain and operations management, ordering, inventory, replenishment, forecasting, transportation, and capacity decisions are distributed across multiple actors and propagate through material, information, and decision flows \citep{jannelli_agentic_2025, trifone2026}. Consequently, understanding autonomy in agentic systems requires examining how decision execution is architecturally distributed across interacting entities \citep{feng_levels_2025, ren2025}.

Rather than converging toward a single architectural model, the surveyed literature reveals recurring organisational principles that characterise distributed decision execution. Based on this synthesis, we identify four recurring architectural patterns according to (i) how decision authority is distributed, (ii) how coordination is achieved, and (iii) how human participation is incorporated throughout execution. Table~\ref{tab:agentic_architectural_paradigms} summarizes these patterns.

\begin{table}[ht] 
\tbl{Synthesis of architectural patterns for distributed decision execution in agentic systems}
{\begin{tabular}{p{2.8cm} p{3.5cm} p{3.5cm} p{2.8cm}} 
\toprule 
\textbf{Architectural Pattern} & 
\textbf{Decision Authority Distribution} & 
\textbf{Primary Coordination Mechanism} & 
\textbf{Representative Literature} \\ 
\toprule
Planner-orientated execution & Centralised planning with delegated execution & Planner-executor coordination, task sequencing& \cite{he2026} \\ 

\midrule

Hierarchical multi-agent execution & Hierarchical delegation across orchestrators and specialised agents & Task allocation, orchestration, supervision & \cite{adam2024,abou2025} \\

\midrule

Collaborative multi-agent execution & Distributed authority among autonomous agents & Negotiation, consensus, information sharing & \cite{jannelli_agentic_2025,maitra2026} \\ 

\midrule

Human-agent collaborative execution & Shared authority between humans and artificial agents & Human oversight, intervention, approval, feedback & \cite{kumar2026} \\ 
\bottomrule 
\end{tabular}}
\label{tab:agentic_architectural_paradigms} 
\end{table}

Table~\ref{tab:agentic_architectural_paradigms} should not be interpreted as a taxonomy of agentic systems. Instead, it represents a synthesis of recurring organizational principles identified across the reviewed literature. While individual studies differ considerably in their implementation details, they consistently organize decision execution by distributing responsibilities across multiple interacting entities \citep{jannelli_agentic_2025}.

Contemporary software frameworks such as CrewAI, AutoGen, LangGraph, Semantic Kernel, and the OpenAI Agents SDK illustrate how these organisational principles are implemented in practice. Although they adopt different software abstractions, including role-based orchestration, conversational coordination, graph-based execution, planner-oriented workflows, and modular agent runtimes, they all operationalize distributed execution by coordinating interactions among planners, orchestrators, specialized agents, external tools, and human actors \citep{vaidhyanathan2025, ren2026}.

Recent supply chain and operations management studies reflect this architectural diversity. Some studies examine collaborative multi-agent systems for autonomous consensus-seeking and negotiation, aiming to improve coordination and reduce phenomena such as the bullwhip effect \citep{jannelli_agentic_2025}. Others investigate multi-agent execution architectures in which agents operate locally while remaining aligned with global constraints or shared operational objectives \citep{maitra2026}. In parallel, governance-oriented studies examine how autonomy, accountability, oversight, and human control should be incorporated into agentic decision-making processes \citep{feng_levels_2025,kumar2026,trifone2026,ren2025}. Table~\ref{tab:positioning_autonomy_coordination} positions representative studies according to their architectural perspective, treatment of autonomy, coordination focus, and system-level outcomes.

\begin{table}[ht]
\tbl{Positioning of this study in relation to agentic AI, autonomy, and coordination literature}{
\begin{tabular}{p{1.9cm} p{2.6cm} p{2.6cm} p{2.6cm} p{2.6cm}}
\toprule
\textbf{Study} &
\textbf{Architectural perspective} &
\textbf{Treatment of autonomy} &
\textbf{Coordination focus} &
\textbf{System-level outcomes} \\
\toprule

\citet{jannelli_agentic_2025} &
Collaborative multi-agent architecture &
Experimental variable: collaboration levels &
Explicit: consensus and negotiation &
Bullwhip effect, global costs, service level \\

\midrule

\citet{maitra2026} &
Multi-agent execution architecture &
Experimental variable: autonomous agents vs. heuristics &
Explicit: shared resource coordination &
Profit, service level, warehouse utilisation \\

\midrule

\citet{kumar2026} &
Human-agent /multi-agent governance architecture&
Governance concept: autonomy trade-offs &
Explicit: orchestration and collaboration phases &
Scalability, fragility, labour disruption \\

\midrule

\citet{trifone2026} &
Human-agent governance architecture &
Governance concept: roles and accountability &
Explicit: coordination effectiveness &
Time/cost efficiency, process visibility \\

\midrule

\citet{ren2025} &
Orchestrated multi-agent architecture &
Governance concept: accountability in manufacturing &
Explicit: subsystem orchestration &
System-level orchestration, efficiency \\

\midrule

\citet{el2026} &
Planner-based agentic optimisation architecture &
Fixed: autonomy embedded in MILP workflow &
Explicit: agent plus expert feedback &
Cycle time, throughput, blocked time \\

\midrule

\citet{parthasarathy2026} &
Human-agent planning architecture &
Assumed: functional autonomy &
Explicit: human-AI collaboration &
Forecast accuracy, planning cycle time \\

\midrule

\textbf{This study} &
\textbf{Distributed decision architecture} &
\textbf{Experimental variable: interaction-based autonomy configurations} &
\textbf{Explicit: coordination under different autonomy settings} &
\textbf{Ordering behaviour, inventory, backlog, bullwhip, system performance} \\

\bottomrule
\end{tabular}}
\label{tab:positioning_autonomy_coordination}
\end{table}

Table~\ref{tab:positioning_autonomy_coordination} highlights three patterns. First, the literature is fragmented across different architectural perspectives, ranging from collaborative multi-agent systems and planner-based optimization workflows to human-agent governance architectures. Second, autonomy is treated inconsistently across research streams. Governance-oriented studies conceptualize autonomy in terms of accountability, oversight, and responsibility allocation, while experimental studies tend to embed autonomy within predefined configurations or compare autonomous agents against benchmark policies. Third, although coordination is frequently examined through negotiation, orchestration, collaboration, and resource allocation mechanisms, limited attention has been given to how specific interaction mechanisms distribute decision authority under different autonomy configurations.

This limitation is central to the present study. If agentic architectures distribute decision execution across multiple entities, then autonomy cannot be assessed only as an individual agent property or as a fixed system-level characteristic. Instead, autonomy must be analysed through the interactions that connect humans, agents, tools, and organizational systems during task planning and execution. These interactions determine whether authority is delegated, shared, constrained, or subject to consultation.

\section{Agentic AI Autonomy Assessment}
\label{sec:aaaa_framework}

\subsection{Operationalising Autonomy through Task Interactions}

The previous section established that autonomy in agentic systems emerges from the allocation of decision authority, the decomposition of goals into executable tasks, and the coordination of distributed decision-making entities. While these concepts provide the theoretical foundations for understanding autonomy, they do not directly provide a mechanism for measuring it. The objective of the proposed AAAA framework is therefore to operationalise autonomy through observable task-level interactions that occur during system execution.

\begin{figure}[th]
    \centering
    \begin{tikzpicture}[
    >=Latex,
    every node/.style={font=\sffamily},
    node distance=2mm and 16mm,
    box/.style={
        draw=gray!30,
        fill=gray!5,
        minimum width=32mm,
        minimum height=12mm,
        rounded corners=4mm,
        inner sep=4mm,
    },
    smallbox/.style={
        box, 
        draw=black!70,
        inner sep=2mm,
        fill=blue!20
    },
    smallsmallbox/.style={
        box, 
        draw=black!70,
        minimum width=18mm, 
        minimum height=8mm,
        inner sep=2mm,
        fill=blue!10
    },
    goalbox/.style={
        smallbox, 
        minimum width=48mm,
        fill=purple!20,
    },
    subgoalbox/.style={
        goalbox, 
        fill=purple!10, 
    },
    actionbox/.style={
        goalbox, 
        fill=teal!20,
    },
    envbox/.style={
        smallbox, 
        fill=cyan!10,
    },
    ]
    
    \pgfdeclarelayer{bgfill}
    \pgfdeclarelayer{background}
    \pgfdeclarelayer{main}
    \pgfdeclarelayer{foreground}
    \pgfsetlayers{bgfill,background,main,foreground}

    % -----------------------
    % AGENT CONTENT
    % -----------------------
    \begin{scope}[local bounding box=maincontent]
    
    \begin{pgfonlayer}{foreground}
        \node[anchor=west] (agent_title) {\textbf{Agentic System}};
        
        % Inner boxes placed BELOW title (no overlap possible)
        \node[smallbox, minimum width=48mm, below=4mm of agent_title] (agent_state) {State};
        % Goal row
        \node[goalbox, below=8mm of agent_state] (agent_goal) {};
        
        \node[anchor=east] at ([xshift=8mm]agent_goal.west) (goal_num) {1};
        \node[anchor=west] at ([xshift=2mm]goal_num.east) (goal_text) {\textbf{Task}};
    \end{pgfonlayer}
    
    % Subgoal row (behind + overlapping)
    \begin{pgfonlayer}{main}
    \node[
        subgoalbox, 
        inner sep=3mm,
        below=-7mm of agent_goal,
        text height=16mm,
        ] (agent_subgoal) {};
    
    \node[anchor=east] at ([xshift=8mm]agent_subgoal.west) (subgoal1no) {1.1};
    \node[anchor=west] at ([xshift=2mm]subgoal1no.east) (subgoal1) {Sub-Task};
    
    \node[below=0mm of subgoal1no] (subgoal2no) {1.2};
    \node[anchor=west] at ([xshift=2mm]subgoal2no.east) (subgoal2) {Sub-Task};

    \node[actionbox, below=8mm of agent_subgoal.south] (action) {Action};
    
    \end{pgfonlayer}
    
    \end{scope}
    
    % -----------------------
    % AGENT WRAP BOX
    % -----------------------
    \begin{pgfonlayer}{bgfill}
    \node[box, fit=(maincontent)] (main) {};
    \end{pgfonlayer}

    % -----------------------
    % User Interactions
    % -----------------------
    \begin{scope}[local bounding box=interactions]
        \node[
            anchor=north east, 
            ] (input_title) at ($(main.north west)+(-12mm,-4mm)$) {\textbf{User Interactions}};
    
        \node[smallbox, left=of agent_subgoal] (L2) {Consultation}; % center-aligned with main
        \node[smallbox,     above=of L2] (L1) {Delegation};
        \node[smallbox,     below=of L2] (L3) {Collaboration};
        \node[smallsmallbox,below=of L3] (L4) {...};
    \end{scope}

    \begin{pgfonlayer}{bgfill}
        \node[box, fit=(interactions)] (interactions) {};
    \end{pgfonlayer}

    % -----------------------
    % Environment
    % -----------------------
    % \begin{scope}[local bounding box=env]
    %     \node[
    %         anchor=north west,
    %     ] (env_title) at ($(main.north east)+(+12mm,-4mm)$) {\textbf{Environment}};
    %     \node[smallbox, right=12mm of agent_subgoal.east] (R1) {Environment};
    % \end{scope}
    
    % \begin{pgfonlayer}{bgfill}
    %     \node[box, fit=(env)] (env) {};
    % \end{pgfonlayer}

    % -----------------------
    % Right box
    % -----------------------
    \node[envbox, right=12mm of agent_subgoal] (env) {Environment};

    % ----------------------
    % Coordinates
    % ----------------------
    \def\xoffset{8mm}

    \coordinate (c1) at ($(agent_subgoal.west)-(\xoffset,-4mm)$);
    \coordinate (c2) at ($(agent_subgoal.west)-(\xoffset,0)$);
    \coordinate (c3) at ($(agent_subgoal.west)-(\xoffset,4mm)$);

    \coordinate (c4) at ($(agent_subgoal.east)+(\xoffset,0)$);
    
    % ----------------------
    % Arrows
    % ----------------------

    % V1 - One to one
    % \draw[thick, ->] (L1.east) -| (c1) -- ($(agent_subgoal.west) + (0,4mm)$);
    % \draw[thick, ->] (agent_subgoal.west) -- (L2.east);
    % \draw[thick, ->] (L3.east) -| (c3) -- ($(agent_subgoal.west) - (0,4mm)$);;

    % V2 - Back and forth, onto main
    % \draw[->] 
    %       ($(interactions.east |- agent_subgoal.west)+(0,2mm)$) --
    %       ($(main.west |- agent_subgoal.west)+(0,2mm)$);
    % \draw[thick, ->]
    %      ($(main.west |- agent_subgoal.west)-(0,2mm)$) --   
    %     ($(interactions.east |- agent_subgoal.west)-(0,2mm)$);

    % V2 - Back and forth, onto subgoal
    \draw[thick, ->] 
          ($(interactions.east |- agent_subgoal.west)+(0,2mm)$) --
          ($(agent_subgoal.west)+(0,2mm)$);
    \draw[thick, ->]
         ($(agent_subgoal.west)-(0,2mm)$) --   
        ($(interactions.east |- agent_subgoal.west)-(0,2mm)$);

    \draw[thick, ->] (agent_state.south) -- (agent_goal.north);
    \draw[thick, ->] (agent_subgoal.south) -- (action.north);

    \draw[thick, dashed, ->] (action.east) -| (env.south);
    %\draw[thick, ->] (env.west) -- (agent_subgoal.east);
    \draw[thick, dashed, ->] (env.north) |- (agent_state.east);
    
\end{tikzpicture}
    \caption{Agentic AI  Autonomy Assessment (AAAA).}
    \label{fig:aaaa}
\end{figure}
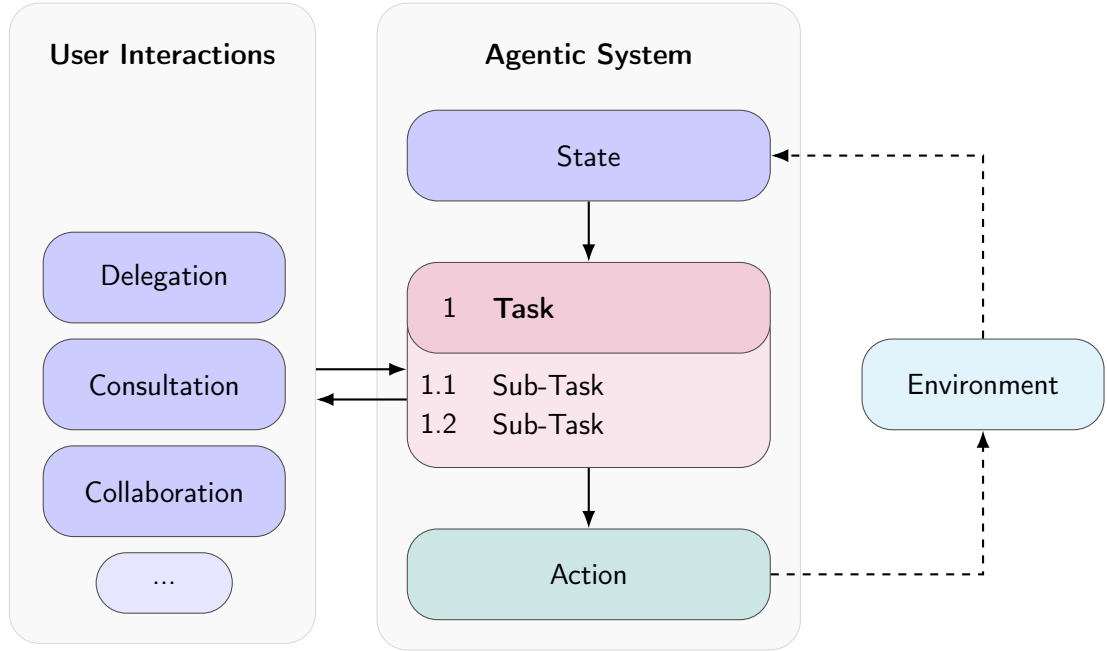

Figure~\ref{fig:aaaa} presents the conceptual organisation of the proposed framework. The framework is organised around three interacting components: the agentic system, external decision-making entities, and the operational environment. Together, these components form a continuous perception-reasoning-action loop in which goals are generated, decomposed into executable tasks, executed, and continuously updated according to changes in the surrounding environment. This execution cycle provides the operational context through which autonomy can be objectively observed and measured.

A central design principle of the proposed framework is that autonomy should be assessed at the level of task execution rather than at the level of individual agents. Contemporary agentic systems rarely perform isolated decisions. Instead, they transform high-level objectives into collections of executable tasks through planning and hierarchical goal decomposition. Consequently, tasks constitute the smallest operational unit through which autonomous behaviour can be consistently observed, compared, and quantified across heterogeneous agentic systems, application domains, and implementation technologies.

Adopting a task-centric perspective shifts the analytical focus from the internal capabilities of intelligent agents towards the interactions established during task execution. Every executable task may involve multiple decision-making entities, including humans, artificial agents, external software services, or organisational systems. These entities contribute to task execution by transferring responsibility, providing additional knowledge, validating intermediate decisions, or jointly participating in planning and execution. Rather than viewing autonomy as an intrinsic property of an agent, the proposed framework evaluates autonomy through these observable task-level interactions. Because task interactions are recorded at the lowest operational level, the resulting measurements can be systematically aggregated to characterise the autonomy of individual agents, multi-agent teams, or entire agentic systems.

Building upon the literature synthesised in Section~\ref{sec:background}, the framework operationalises task interactions through three recurring interaction patterns that consistently emerge across distributed decision-making environments:

\begin{itemize}

\item \textbf{Delegation.}
A decision-making entity transfers responsibility for executing a task to another entity while expecting autonomous completion within predefined operational boundaries.

\item \textbf{Consultation.}
A decision-making entity requests additional knowledge, validation, preferences, or approval from another entity before continuing task execution while retaining overall responsibility for the decision.

\item \textbf{Collaboration.}
Two or more decision-making entities jointly contribute to planning, refining, or executing the same task through shared decision-making and mutual coordination.

\end{itemize}

These interaction patterns should not be interpreted as an exhaustive taxonomy of human-AI interaction. Instead, they represent three recurring forms of task-level contribution that are sufficiently general to characterise distributed decision-making while remaining operationally measurable. Additional interaction patterns may be incorporated for domain-specific applications, provided that their contribution to task execution can be explicitly defined and objectively quantified.

The proposed framework is compatible with the execution principles adopted by contemporary agentic AI architectures. Frameworks such as ReAct operationalise complex objectives through iterative reasoning and acting cycles, progressively decomposing goals into executable subtasks during execution \citep{yao_react_2023}. More recent architectures, including ReAcTree, further extend this principle by distributing goal decomposition across multiple specialised agents responsible for different branches of the execution hierarchy. Similar hierarchical task-management strategies are increasingly adopted by modern agentic systems such as Claude Code and related software engineering agents, where complex objectives are coordinated through dynamically generated task structures rather than fixed execution workflows \citep{anthropic_claude_2026, liu_dive_2026}.

Although these architectures differ substantially in their implementation, they share a common execution principle: operational objectives are realised through the coordinated execution of hierarchically decomposed tasks. This common characteristic aligns directly with the task-centric perspective adopted by the proposed framework. By selecting executable tasks as the primary unit of analysis, the framework remains independent of specific software implementations while remaining applicable across a broad range of contemporary agentic architectures.

Tracking task interactions throughout execution enables autonomy to be quantified objectively through observable operational behaviour rather than inferred from software architectures or internal reasoning processes. Because task interactions can be automatically recorded during execution, the proposed framework supports continuous autonomy assessment throughout the lifecycle of an agentic system. At the operational level, the resulting metrics enable comparisons across tasks, agents, and heterogeneous agentic systems. At the organisational level, continuous monitoring provides opportunities to identify behavioural drift, evolving interaction patterns, and unintended changes in operational autonomy over time \citep{chen_ai_2026}.

\subsection{Structural Components of the Framework}

The proposed AAAA is composed of a set of structural components that collectively describe how autonomy is operationalised during task execution. These components define the conceptual organisation of the framework by representing the internal decision-making process of the agentic system, its interactions with external entities, and its relationship with the surrounding environment. Rather than corresponding to specific software modules, they provide an implementation-independent abstraction that can be applied across different agentic architectures.

Table~\ref{tab:framework_components} summarises the structural components adopted by the proposed framework together with their respective roles in autonomy assessment.

\begin{table}[ht]
\tbl{Structural components of the proposed autonomy assessment framework.}{
\begin{tabular}{p{2cm} p{6cm} p{5cm}}
\hline
\textbf{Component} &
\textbf{Role in the framework} &
\textbf{Contribution to autonomy assessment} \\
\hline

State &
Represents the internal knowledge of the agentic system, including memory, contextual information, operational constraints, and the current representation of the environment. &
Provides the informational context from which tasks are generated and updated throughout execution. \\

Goal &
Represents a desired outcome pursued by the agentic system. Goals may originate from external delegation or emerge internally during execution. &
Guides the intended behaviour of the system and initiates the task generation process. \\

Task &
Represents an executable unit derived from one or more goals through planning and goal decomposition. &
Serves as the primary unit of analysis through which autonomy and user involvement are measured. \\

Action &
Represents the operational execution of one or more tasks within the surrounding environment. &
Produces the observable behaviour resulting from autonomous decision-making. \\

External Interactions &
Represent contributions from humans, other agents, external software services, or organisational systems throughout task execution. &
Provide the observable interactions used to quantify autonomy through delegation, consultation, and collaboration. \\

Environment &
Represents the operational context in which the agentic system senses information and executes actions. &
Provides continuous feedback that updates the agent state and influences subsequent decisions. \\

\hline
\end{tabular}}
\label{tab:framework_components}
\end{table}

The proposed framework is centred around the concept of the agent state, which represents the current knowledge available to the agentic system. Consistent with classical Belief-Desire-Intention (BDI) architectures \citep{georgeff_belief-desire-intention_1999}, the state aggregates contextual information, operational constraints, memory, and environmental observations that collectively define the information upon which subsequent decisions are based. Unlike implementation-specific memory structures, the state is treated here as a conceptual representation of the knowledge available to the agent at any point during execution.

Based on its current state and external interactions, the agentic system generates one or more operational goals representing desired future states. Goals may originate directly from external delegation or emerge internally as intermediate objectives generated during planning and reasoning. While goals describe what the system intends to accomplish, they are not directly executable. Contemporary agentic systems therefore operationalise goals through planning and hierarchical goal decomposition, transforming high-level objectives into collections of executable tasks.

Within the proposed framework, tasks constitute the central analytical construct. Although goals define the intended behaviour of the agentic system, tasks represent the operational realisation of these objectives and therefore provide the smallest observable unit through which autonomous behaviour can be assessed. Every task possesses a clear operational purpose, may involve one or multiple decision-making entities, and ultimately results in one or more executable actions. 

Actions represent the execution of tasks within the operational environment. Once executed, actions modify the surrounding environment, generating new observations that update the agent state and potentially trigger the creation of new goals and tasks. This continuous feedback establishes the perception, reasoning and action cycle characteristic of contemporary agentic systems while providing the temporal context in which autonomy evolves throughout execution.

Finally, the proposed framework explicitly incorporates external interactions as structural components rather than treating them as external events. During task execution, humans, artificial agents, software services, and organisational systems may contribute to task completion by transferring responsibilities, providing additional knowledge, validating intermediate decisions, or jointly participating in planning and execution. These observable contributions constitute the basis for the interaction patterns introduced in the following subsection and ultimately provide the information required to quantify autonomy.

\subsection{Task Interaction Mechanisms}

The structural components introduced in the previous subsection describe how an agentic system generates goals, decomposes them into executable tasks, and interacts with its operational environment. However, autonomy is not determined solely by these internal processes. Throughout task execution, external decision-making entities may influence how tasks are initiated, planned, executed, or validated. Consequently, autonomy should be understood not only as an intrinsic capability of the agentic system, but also as a property that emerges from the distribution of responsibilities between the agentic system and external entities during task execution.

This observation motivates the central premise of the proposed framework: task interactions constitute the observable events through which operational autonomy can be assessed. Rather than characterising interactions according to software-specific communication protocols or implementation details, the framework classifies interactions according to their effect on task execution. Specifically, interactions are distinguished by how they modify three fundamental task properties: task ownership, decision authority, and execution responsibility.

\begin{figure}[th]
    \centering
    \includegraphics[width=\textwidth]{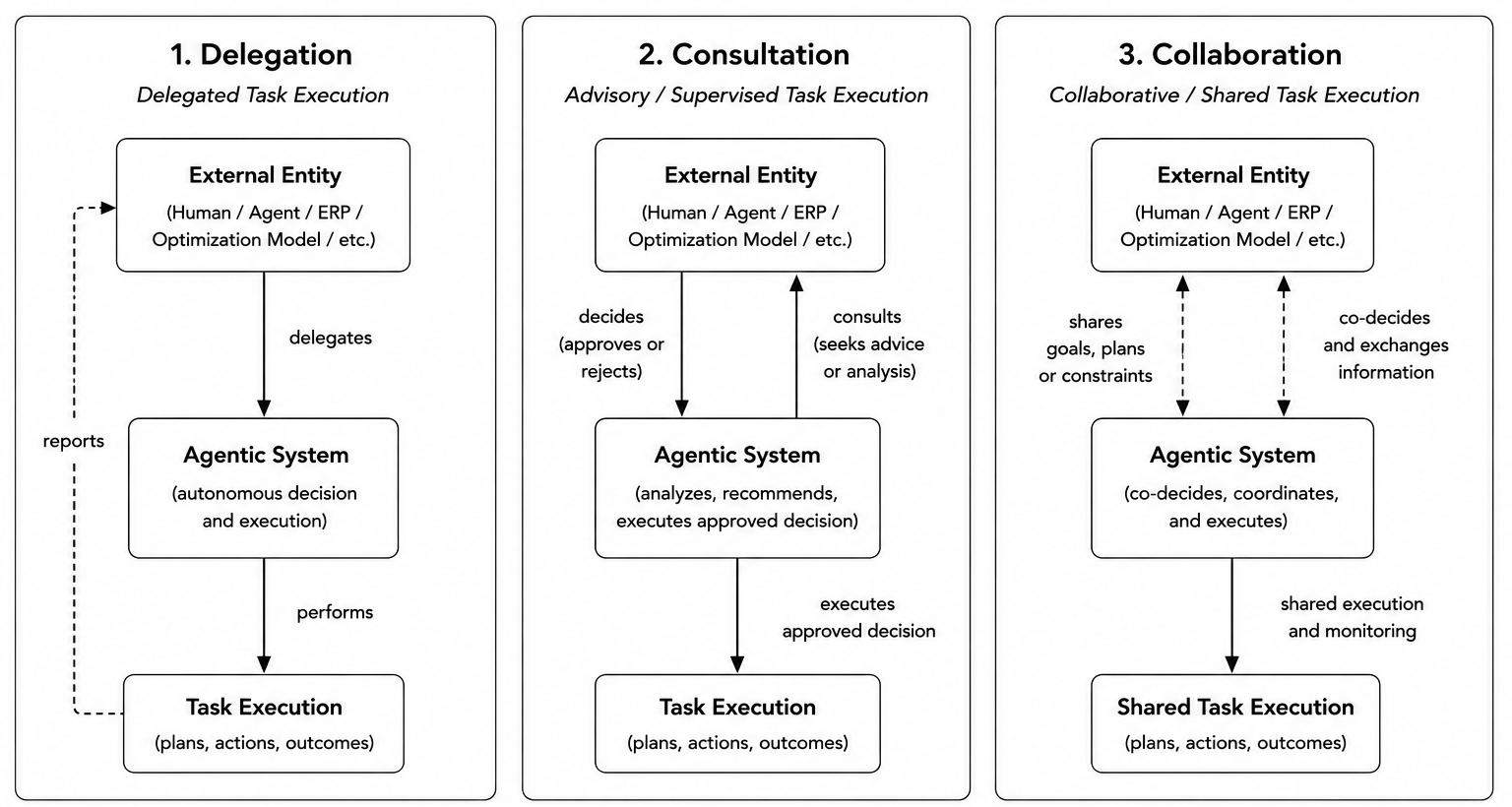}
    \caption{Task interaction mechanisms considered by the proposed framework. The three interaction mechanisms represent alternative distributions of responsibility over the same executable task. Solid arrows denote the primary execution flow, whereas dashed arrows represent information exchange and coordination interactions.}
    \label{fig:interaction_patterns}
\end{figure}

Figure~\ref{fig:interaction_patterns} illustrates the three interaction mechanisms considered in the proposed framework together with their effect on these properties. Although the interaction mechanisms differ operationally, they all describe alternative distributions of responsibility over the same executable task rather than different categories of agentic systems. The proposed interaction mechanisms should not be interpreted as communication or interaction protocols. Instead, they represent categories of task-level responsibility allocation that are independent of the underlying communication protocol or implementation technology.

The three interaction mechanisms illustrated in Figure~\ref{fig:interaction_patterns} should be understood as alternative configurations for distributing responsibilities during task execution. Rather than defining different categories of agentic systems, they describe how task ownership, decision authority, and execution responsibility may be allocated between an external decision-making entity and an agentic system. Consequently, the proposed framework evaluates autonomy through the distribution of these responsibilities rather than through the internal capabilities of the agentic system itself.

From the perspective of the proposed framework, delegation represents the principal mechanism through which autonomous initiative becomes observable. Because task ownership and decision authority are transferred to the agentic system, delegated tasks reveal the extent to which the system can independently transform externally assigned objectives into executable actions. Delegation therefore operationalises the origin of task execution and provides the empirical basis for measuring initiative throughout the execution process.

In contrast, consultation captures situations in which autonomous execution becomes conditional upon external intervention. Although the agentic system remains responsible for performing the operational task, critical decision points require additional information, preferences, validation, or approval from an external entity before execution can proceed. Consultation therefore does not reduce the system's execution capability; instead, it reveals where decision authority remains intentionally external. Consequently, consultation operationalises decision control and provides observable evidence of the degree of external oversight exercised during task execution. For instance, a consultation interaction may be implemented through synchronous API calls, message-passing protocols, or LLM tool invocations. Although these protocols differ technically, they all correspond to the same observable consultation event within the proposed framework.

Finally, collaboration represents situations in which task ownership, decision authority, and execution responsibility are intentionally shared among multiple decision-making entities. Rather than transferring responsibility or introducing isolated interventions, collaboration reflects continuous joint participation throughout planning and execution. Although collaborative interactions are not directly incorporated into the quantitative autonomy metrics proposed in this study, they constitute an extensible interaction mechanism that enables the framework to represent richer forms of distributed decision-making and shared autonomy frequently observed in multi-agent and human-AI systems.

Collectively, these three interaction mechanisms establish the observable events upon which the proposed autonomy assessment framework is built. Because each interaction modifies a different aspect of task execution, they provide an implementation-independent representation of external involvement that can subsequently be translated into quantitative autonomy metrics. 

\subsection{Operational Autonomy Metrics}

The interaction mechanisms introduced in the previous subsection establish the observable events through which autonomy can be assessed during task execution. Building upon these interactions, the proposed framework operationalises autonomy through two complementary dimensions: autonomous initiative and decision control. Autonomous initiative reflects the extent to which executable tasks are generated internally by the agentic system, whereas decision control reflects the extent to which task execution remains dependent on external decision-making entities. 

The first dimension, referred to as autonomous initiative, captures the ability of an agentic system to generate executable tasks through planning and goal decomposition. Since task generation is directly observable through delegation interactions, autonomous initiative is measured by comparing the number of internally generated tasks to the total number of executable tasks.
The resulting Initiative Rate (IR) is defined as

\begin{equation}
\label{eq:initiative_rate}
IR=\frac{N_I}{N},
\end{equation}

where

\begin{align*}
N &= \text{Total number of executable tasks},\\
N_I &= \text{Number of internally generated tasks}.
\end{align*}

Values approaching zero indicate that task execution is almost entirely determined by externally delegated tasks, whereas values approaching one indicate that a substantial proportion of executable tasks emerge through autonomous planning and hierarchical goal decomposition.

The second dimension, referred to as decision control, evaluates the extent to which task execution depends upon external decision-making after a task has been delegated. Although an agentic system may autonomously perform planning and execution, consultations requesting additional knowledge, approval, operational preferences, or validation reveal situations in which decision authority intentionally remains external. Decision control is therefore operationalised through observable consultation events occurring during task execution.

The Consultation Rate (CR) is defined as

\begin{equation}
\label{eq:consultation_rate}
CR=\frac{N_C}{N},
\end{equation}

where

\begin{align*}
N &= \text{Total number of executable tasks},\\
N_C &= \text{Number of tasks requiring consultation}.
\end{align*}

Lower consultation rates indicate that tasks are completed with limited external intervention, whereas higher consultation rates reveal increasing dependence on external decision-making entities throughout execution.

Because autonomous initiative and decision control characterize complementary aspects of operational autonomy, both dimensions are normalized to the interval $[0,1]$ and combined into a composite autonomy indicator. The resulting Autonomy Score (AS) is defined as

\begin{equation}
\label{eq:autonomy_score}
AS=w_1IR+w_2(1-CR),
\end{equation}

subject to

\begin{align*}
w_1 &\geq 0,\\
w_2 &\geq 0,\\
w_1+w_2 &=1.
\end{align*}

The weighting coefficients allow the framework to accommodate different operational contexts while preserving a common measurement structure. Environments emphasising independent task generation may assign greater importance to autonomous initiative, whereas safety-critical or highly regulated environments may place greater emphasis on maintaining decision authority through external control mechanisms.

It is important to emphasise that the proposed Autonomy Score should not be interpreted as a measure of intelligence, reasoning quality, or decision performance. Instead, it quantifies how operational responsibilities are distributed between an agentic system and external decision-making entities throughout task execution. By grounding the metric in observable task interactions rather than internal implementation details, the framework provides a technology-independent mechanism for comparing autonomy across different agentic systems, software frameworks, and operational contexts. 
It can be applied during development to calibrate the system through iterative testing on a range of tasks and adjusting of the system parameters until the desired autonomy range is reached. Due to the stochasticity of LLM-based systems, the unpredictability of the execution environment and the emerging of new tasks, the calibrated autonomy range cannot be guaranteed during runtime. Therefore, the framework should be implemented directly into the runtime execution to monitor how the degree of autonomy changes over time and for different tasks.

\section{Experimental Setup}
\label{sec:validation}

In this section the AAAA is applied to an agentic supply chain system to study the effects of the autonomy score on a well-known simulated supply chain context, providing a controlled environment for demonstration purposes.

\subsection{The Beer Distribution Game}
The beer distribution game is a well-known supply chain simulation in which players try to minimise their companies' costs through appropriate inventory management under demand uncertainty. The interactions of different supply chain tiers, their feedback, and added time delays create dynamics such as the bullwhip effect \citep{lee_bullwhip_1997}. These emerging dynamics have been addressed in a variety of studies, providing a well-researched experimental context for investigating end-to-end supply chains \citep{sterman_modeling_1989, sterman_business_2000}.

\begin{figure}[ht]
    \centering
    \begin{tikzpicture}[
    >=Latex,
    scale=0.8,
    every node/.style={font=\sffamily, transform shape},
    node distance=8mm and 7mm,
    box/.style={
        draw=gray!40,
        fill=gray!5,
        minimum width=20mm,
        minimum height=10mm,
        rounded corners=4mm,
        inner sep=3mm,
    },
    supplybox/.style={
        draw=black!70,
        fill=teal!20,
        minimum height=8mm,
        trapezium,
        trapezium left angle=70,
        trapezium right angle=110,
        thick
    },
    demandbox/.style={
        draw=black!70,
        fill=cyan!10,
        ellipse,
        minimum height=8mm,
        thick
    },
    tierbox/.style={
        box,
        draw=black!70,
        fill=blue!20
    }
]

% -----------------------
% Nodes
% -----------------------

\node[supplybox] (supply) {Supply};

\node[tierbox, right=of supply] (factory) {Factory};

\node[tierbox, right=of factory] (distributor) {Distributor};

\node[tierbox, right=of distributor] (wholesaler) {Wholesaler};

\node[tierbox, right=of wholesaler] (retailer) {Retailer};

\node[demandbox, right=of retailer] (customer) {Demand};

% -----------------------
% Arrows
% -----------------------

\draw[thick, ->] (supply) -- (factory);
\draw[thick, ->] (factory) -- (distributor);
\draw[thick, ->] (distributor) -- (wholesaler);
\draw[thick, ->] (wholesaler) -- (retailer);
\draw[thick, ->] (retailer) -- (customer);

\end{tikzpicture}
    \caption{Beer Distribution Supply Chain}
    \label{fig:supply_chain}
\end{figure}
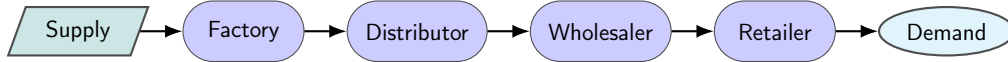

To set up the experimental context, a modified version of the beer distribution game is implemented as a digital supply chain simulation. The supply chain consists of a factory, distributor, wholesaler, and retailer tiers, as shown in Fig. \ref{fig:supply_chain}. Upstream of the factory is the beer supply, which creates new products according to the factories' production orders. Downstream of the retailer sits the customer, who provides the demand pattern.

The parameters for the simulation are described in Table \ref{tab:sim_parameters} and provide a fixed baseline for all experiments. The starting inventory for each tier is set to 16 units. The demand pattern follows a seasonal shape, with two peaks at week 3 and week 9. The total length of each simulation is 10 weeks. The time from an order placement until the order becomes visible at the upstream tier is 2 weeks. After dispatching a shipment, it takes another 2 weeks to reach the downstream tier. The total lead time from order placement until delivery is therefore at a minimum of 4 weeks.

\begin{table}[th]
\tbl{Simulation Parameters}{
\begin{tabular}{p{5cm}p{2cm}}
\toprule
\textbf{Simulation Parameter} &
\textbf{Value} \\
\toprule
Starting Inventory & 16 units\\
Demand Pattern & [4, 8, 12, 8, 4, 2, 4, 8, 12, 8] \\
Simulation Length & 10 weeks \\
Information Lead Time & 2 weeks \\
Shipping Lead Time & 2 weeks \\
Holding Cost & 1€/unit \\
Backlog Cost & 2€/unit \\
\bottomrule
\end{tabular}}
\label{tab:sim_parameters}
\end{table}

\subsection{Agentic Architecture Workflow}

The agentic workflow is implemented using LangGraph. The simulation of the game is running on a local development server that exposes a REST API with endpoints to manage the simulation workflow (setting up a new game, advancing the week, querying state) and endpoints for company interactions (creating shipments, placing orders).

\begin{figure}
\centering
\subfloat[Game Graph]{% 
\includegraphics[scale=0.45]{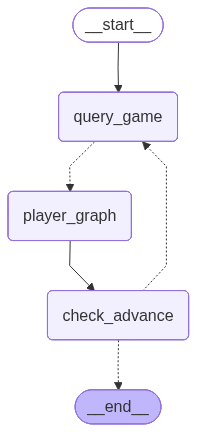}}\hspace{20pt}
\subfloat[Player Subgraph]{%
\includegraphics[scale=0.45]{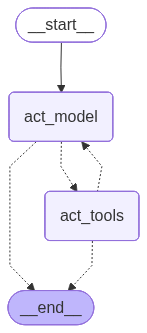}}
\caption{Agentic Workflow Graph Implmented using LangGraph}
\label{fig:game_player_graphs}
\end{figure}

The agentic workflow is implemented as a graph, consisting of a main game graph that controls the overall simulation flow and separate sub-graphs for each tier. Fig. \ref{fig:game_player_graphs} visualises the nodes and edges contained in the graph. The entry of the game graph is the query\_game node, which queries the current game state from the simulation server and updates the local state, such as the current week and simulation status. Next, this node fans out into four player graphs that execute in parallel. After every player finishes their subgraph, the game graph continues to the check\_advance node, which resets part of the state for the coming week and advances the week on the server. After, the graph either loops around if the status is still running or ends execution if the server returns that the simulation is complete.

% \subsection{Agentic AI as Supply Chain Tiers}
The agentic system is modelled as player sub-graphs, each consisting of the two nodes act\_model and act\_tools, based on the ReAct architecture \citep{yao_react_2023}. The sub-graph architecture further encapsulates a separate state from the main graph and other sub-graphs, ensuring no confidential information is leaked between players. The graph entry is the act\_model node, which updates the player company's state and assembles a game context. Afterwards, the context is passed to a local LLM. The model reasons over the provided context and decides which tools to call from a defined list of available tools, shown in Table~\ref{tab:player_tools}. The response of the LLM is then parsed for any tool calls, which are handed over to the act\_tools node. The act\_tool node is then responsible for handling the execution of the tool call. If the LLM returns a tool call to "done" the subgraph execution ends and control is returned to the game graph. Otherwise, the tool response is appended to the message history and passed back into the act\_model node for the next reasoning step.

\begin{table}[t]
\tbl{Available Player Tools}{
\begin{tabular}{p{2.5cm} p{2.5cm} p{8cm}}
\toprule
\textbf{Tool Name} &
\textbf{Parameters} &
\textbf{Function} \\
\toprule
place\_order & 
\begin{minipage}[t]{\linewidth}
\begin{itemize}[leftmargin=*, noitemsep, topsep=0pt]
    \item \textit{company\_id}: str
    \item \textit{supplier\_id}: str
    \item \textit{quantity}: int
\end{itemize}
\end{minipage}
& Places a replenishment order with the given quantity at the specified supplier company. \\
\midrule
create\_shipment & 
\begin{minipage}[t]{\linewidth}
\begin{itemize}[leftmargin=*, noitemsep, topsep=0pt]
    \item \textit{company\_id}: str
    \item \textit{order\_id}: str
    \item \textit{quantity}: int
\end{itemize}
\end{minipage}
& Creates a shipment against an open order specified by its order\_id. Partial shipments are possible.\\
\midrule
read\_tasks & - & Returns the current list of open tasks.\\
\midrule
create\_task & 
\begin{minipage}[t]{\linewidth}
\begin{itemize}[leftmargin=*, noitemsep, topsep=0pt]
    \item \textit{parent\_task\_id}: str
    \item \textit{content}: str
\end{itemize}
\end{minipage}
& Creates a new task with details specified inside the content parameter. If no parent\_task\_id is supplied this will create a high-level task (e.g. task\_1). For decomposing high-level goals a parent\_task\_id can be supplied, which will create the new task as a subtask (e.g. task\_1\_1). \\
\midrule
updated\_task & 
\begin{minipage}[t]{\linewidth}
\begin{itemize}[leftmargin=*, noitemsep, topsep=0pt]
    \item \textit{task\_id}: str
    \item \textit{new\_status}: \mbox{Literal}['pending', 'in\_progress', 'completed']
\end{itemize}
\end{minipage}
& Used to update the status of a task, to keep track of overall progress. Tasks can be either 'pending', 'in\_progress' or 'completed'.\\
\midrule
remove\_task & 
\begin{minipage}[t]{\linewidth}
\begin{itemize}[leftmargin=*, noitemsep, topsep=0pt]
    \item \textit{task\_id}: str
\end{itemize}
\end{minipage}
& Removes a task from the task store. Only called when a task was created in error or has become irrelevant. \\
\midrule
request\_consultation &
\begin{minipage}[t]{\linewidth}
\begin{itemize}[leftmargin=*, noitemsep, topsep=0pt]
    \item \textit{task\_id}: str
    \item \textit{details}: str
\end{itemize}
\end{minipage}
& Can be used to request consultation from the user entity for a given task.\\
\midrule
done & - & Ends the player subgraph execution \\
\bottomrule
\end{tabular}}
\label{tab:player_tools}
\end{table}

\subsubsection{The Task System}
The task system is operationalised as a task store implemented as a typed dictionary and stored in the game state. Whenever a sub-graph is executed, a reference to the players' task store is passed into the sub-graph, updated by the player as needed, and then stored back into the game state.
The agentic system starts each simulation with three pre-defined goals, which provide high-level guidance and cannot be removed or deleted to encourage the player to perform goal decomposition. Open tasks are always passed into the context, and players can interact with the task system through the provided tools. The default tasks are defined as follows:

\begin{description}
    \item[task\_1]: Minimise your own total cost over time (holding + backlog).
    \item[task\_2]: Contribute to overall system stability by minimising the bullwhip effect.
    \item[task\_3]: Maintain a service level of at least 95\%.
\end{description}

%For the experiments the focus player was assigned all 3 tasks, while the other players got only the first task.
The agentic system can request consultation from the user for a specific task. Consultation is implemented as a binary decision of "approved" or "rejected". To keep the effect of consultations consistent across experiments, the user is modelled as a simple function that always returns approval for a given task. This way, the consultation rate can be tracked accurately while the model actions are not influenced.

\subsubsection{Experiment Variables}
The goal of this experiment is to study the potential relation between the autonomy of an agentic systems and the performance of a supply chain. The supply chain performance is measured using both the costs and fill rate observed per tier. To span the maximum range of the autonomy spectrum, the system prompt of the agentic system is adjusted along 5 levels (L1-L5) to encourage behaviours that would result in different autonomy scores. 
The following is the direct mapping for the system prompt augmentation, derived from the five levels presented in the synthesis of user-agent responsibilities in Section \ref{sec:background}, Fig. \ref{fig:goal_decomposition_autonomy}:

\begin{description}
    \item[L1]: The user should be in charge at all times and drive all decisions and plans.
    \begin{itemize}[leftmargin=*, noitemsep, topsep=0pt]
        \item You do not take action that involves subjective judgement without explicit user instruction.
        \item If you suggest actions, you do not execute them without user approval.
    \end{itemize}
    \item[L2]: The user is your collaborator. Collaboratively plan, delegate, and execute decisions.
    \begin{itemize}[leftmargin=*, noitemsep, topsep=0pt]
        \item Aim to strike a balance between the user and your own agency.
        \item Frequent back-and-forth communication with the user is encouraged.
    \end{itemize}
    \item[L3]: The user acts as a consultant. You take the lead, but consult the user for their expertise or preferences.
    \begin{itemize}[leftmargin=*, noitemsep, topsep=0pt]
        \item Take initiative in task planning and execution.
        \item The user can provide feedback and higher-level directional guidance within the given boundaries.
        \item Carefully consider the topic and timing of the user consultation.
    \end{itemize}
    \item[L4]: The user is an approver. You may only engage the user in risky or pre-specified scenarios.
    \begin{itemize}[leftmargin=*, noitemsep, topsep=0pt]
        \item Only consult the user when you encounter a blocker that you cannot resolve on your own.
        \item E.g. consult after reaching a failure state that prevents workflow continuation, when providing credentials, or signing off on consequential actions.
    \end{itemize}
    \item[L5]: The user is the observer. You operate fully independently, while the user is only monitoring.
    \begin{itemize}[leftmargin=*, noitemsep, topsep=0pt]
        \item If you encounter a blocker, iterate on solutions or modify your approach until resolved without user involvement.
        \item The user cannot provide input but can monitor actions and activity logs.
    \end{itemize}
\end{description}

To achieve a reasonable number of experiments, the prompt is varied across L1...L5 at each tier, while the augmentation for the other tiers is held constant at level L3, aiming for a balanced user-agent interaction. For each tier, the experiment is repeated twelve times, optimising for breadth while keeping the results manually traceable. With a total of 60 runs per tier, the resulting number of experiments was therefore 240.

\subsection{Results}
\begin{figure}[th]
    \centering
    \includegraphics[width=\textwidth]{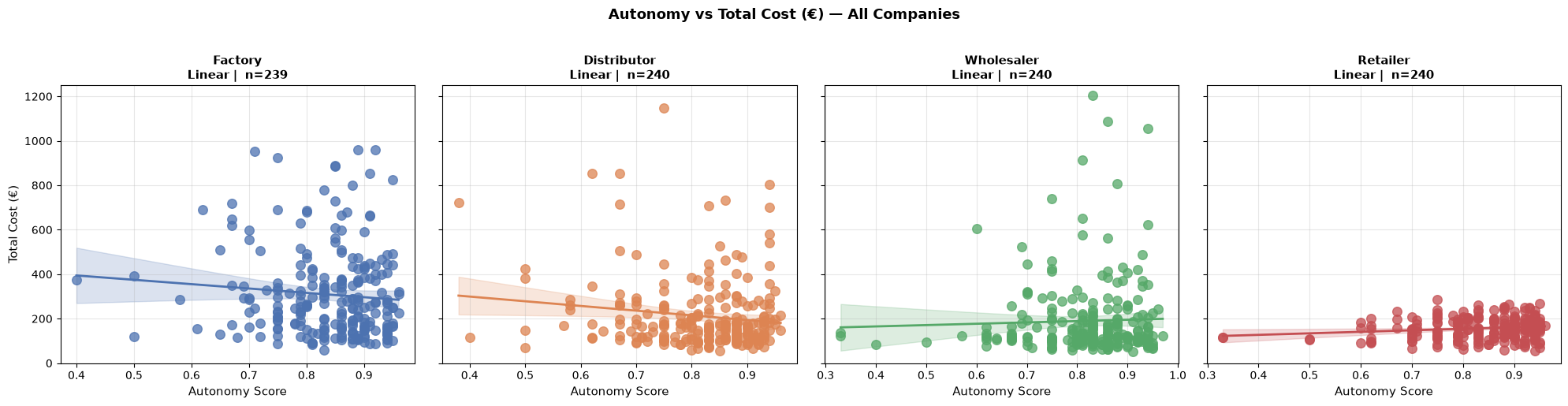}
    \caption{Autonomy-Cost Relationship Across Supply Chain Tiers Indicating a Performance Split Between Upstream and Downstream Actors}
    \label{fig:result_all_cost}
\end{figure}

To analyse the results, the autonomy score is calculated per tier for each experimental run. The data spans the autonomy score range of 0.33-0.97 with a mean of 0.83. The data is skewed towards the left (skewness: -1.465, p<0.0001) with samples clustering in the higher autonomy range and only a few data points in the lower ranges. Table \ref{tab:results} shows the results for cost and fill rate of the experiments separated by supply chain tier. The cost results are visualised in Fig. \ref{fig:result_all_cost} using an ordinary least squares (OLS) regression with a linear fit.

\begin{table}
\setlength{\tabcolsep}{3.7pt}
\tbl{Experimental Results of a Supply Chain Tiers Autonomy and Their Associated Cost and Fill Rate.}
{\begin{tabular}{lccccccccccc} \toprule
 & \multicolumn{4}{l}{Cost} & & \multicolumn{6}{l}{Fill Rate} \\ 
 \cmidrule{2-5} \cmidrule{7-12}
 Tier & Mean & Std & \makecell{OLS\\Coef} & \makecell{OLS\\p-value} & & 
        Mean & Std & \makecell{OLS\\Coef} & \makecell{OLS\\p-value} & \makecell{Tobit\\Coef} & \makecell{Tobit\\p-value} \\ 
 \midrule
 Factory & 308.49€ & 196.16€ & -195.64€ & 0.167 && 82.04\% & 23.45\% & -13.58\% & 0.423 & - & - \\
Distributor & 211.24€ & 157.60€ & -211.14€ & \cellcolor{siggreen}0.028 && 56.75\% & 30.45\% & 16.16\% & 0.388 & 20.80\% & 0.355 \\
Wholesaler & 190.15€ & 169.53€ & 60.04€ & 0.573 && 64.24\% & 25.51\% & -15.27\% & 0.340 & -13.93\% & 0.440 \\
Retailer & 157.11€ & 47.06€ & 68.45€ & \cellcolor{siggreen}0.018 && 29.00\% & 7.22\% & -9.60\% & \cellcolor{siggreen}0.030 & -9.60\% & \cellcolor{siggreen}0.030 \\
\bottomrule
\end{tabular}}
%\tabnote{\textsuperscript{a}This footnote shows how to include footnotes to a table if required.}
\label{tab:results}
\end{table}

From all the tiers, the factory has both the highest cost and the highest variance, with cost and variance decreasing towards the downstream tiers. In the distributor tier, costs and variance are lower, with the same negative slope. The wholesaler shows more clustering towards the lower cost end, with few outliers in the upper autonomy range and a slight positive trend curve. Finally, the retailer has the tightest cluster with data points concentrated at the lower end and little variance. The costs are slightly increasing with higher autonomy. 

The results seem to show an overall trend of a negative autonomy-cost relationship for the upstream actors, while downstream actors seem to experience the opposite effect. This results in lower costs for the factory and distributor, while the wholesaler and retailer incur higher costs at higher autonomy levels. The suggested upstream-downstream split is supported by two significant data anchors (p<0.05) for the distributor and retailer coefficients, while also being affected by noisy data and high variance. The slope of the wholesaler tier being close to zero results in a high p-value of 0.573, which would be consistent with a transition point between upstream and downstream responses.

\begin{figure}[th]
    \centering
    \includegraphics[width=\textwidth]{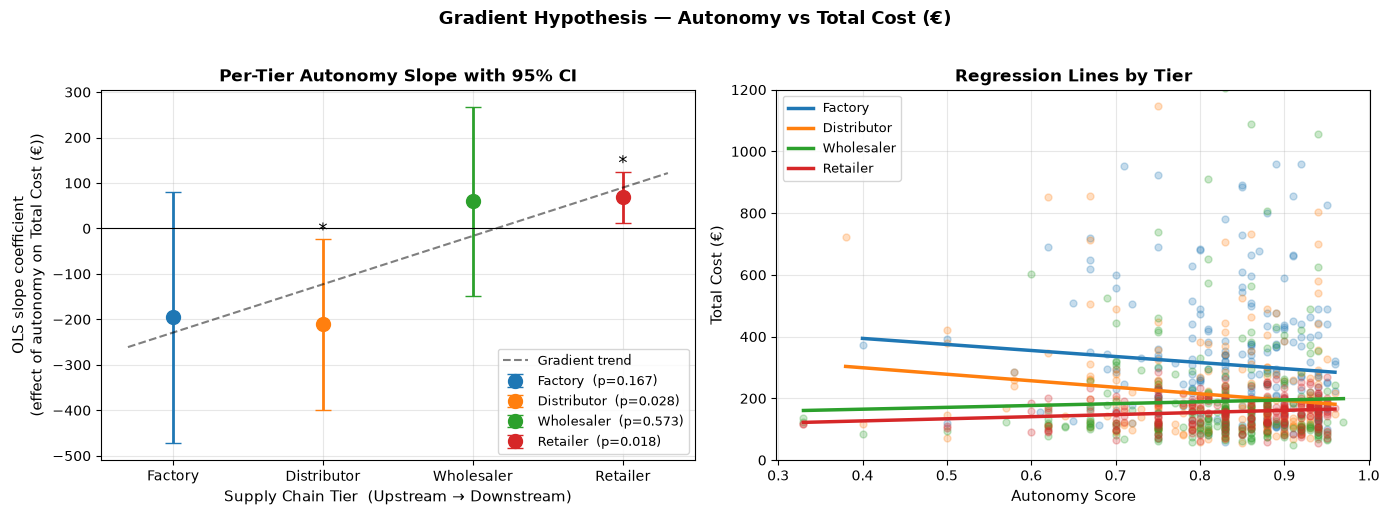}
    \caption{Combined Linear Regression Model for the Relation of Supply Chain Tier with Simulation Cost}
    \label{fig:result_all_gradient}
\end{figure}

% ----- POOLED REGRESSION
To support further analysis of this pattern, the slopes of each tier are compared and combined into a pooled linear regression model, based on all collected data points. The left-hand side of Figure \ref{fig:result_all_gradient} shows the linear regression coefficient for each tier, including an error band. The dashed line shows the trend line for the slopes along the supply chain tiers. The right-hand side of the same figure shows the individual tier regression lines plotted on top of each other. The combined regression results in a positive coefficient ($\beta$=+109.6, p=0.015), indicating that with each supply chain tier, the cost gradient increases. This means upstream tiers financially benefit from higher autonomy, while downstream tiers are getting hurt by it.

\section{Discussion}
\label{sec:Dicussion}

The results provide several important insights into the role of autonomy in agentic supply chain systems, particularly with respect to governance, system-level dynamics, and managerial decision-making.

A central finding of this study is that the autonomy score does not exhibit a uniform or consistently positive relationship with system performance. While autonomy is often implicitly associated with improved efficiency and decision quality in agentic AI literature, the experimental results suggest a more nuanced interpretation. Specifically, the experiment results showed that higher autonomy did not universally translate into improved fill rates or lower operational costs. Instead, aligns with \citet{bradshaw_dimensions_2004}, which argues that autonomy primarily influences how decision-making responsibilities are distributed, rather than directly determining performance outcomes. This distinction is critical, as it positions autonomy as a governance variable rather than a performance variable \citep{chiris_aura_2025}. In this sense, the proposed autonomy score provides a mechanism to monitor and regulate the degree of independence of agentic systems, without assuming that higher autonomy is inherently desirable \citep{kumar_balancing_2026, onnasch_human_2014}.

The experimental results further reveal a position-dependent relationship between autonomy and cost across the supply chain. Upstream actors, particularly the factory and distributor, tend to benefit from higher levels of autonomy, as reflected in decreasing cost trends. In contrast, downstream actors such as the retailer exhibit increasing costs with higher autonomy. This asymmetry suggests that the effectiveness of agentic autonomy is contingent upon the structural position of an actor within the supply chain and should neither be universally desirable \citep{oneill_humanautonomy_2022}, or universally undesirable \citep{mitchell_fully_2025}. One possible explanation lies in the propagation of uncertainty and information distortion, commonly associated with the bullwhip effect. Upstream actors operate under higher levels of uncertainty due to demand amplification and delayed information signals, which may make autonomous goal decomposition and proactive decision-making more beneficial. Conversely, downstream actors are more directly exposed to customer demand and may benefit from tighter human oversight to avoid overreaction or biased decision amplification.
These findings highlight that autonomy should not be treated as a one-size-fits-all design parameter. Instead, optimal autonomy configurations appear to depend on both the decision context and the position within the supply chain. This reinforces the importance of adopting a differentiated approach to agentic system design, where autonomy levels are deliberately calibrated rather than uniformly maximised.

From a governance perspective, the autonomy score provides a structured way to operationalise and monitor these differences. By capturing both autonomous initiative and decision control, the metric enables organisations to define acceptable autonomy boundaries and track deviations over time, enabling oversight, accountability and regulatory compliance \citep{chiris_aura_2025,kumar_balancing_2026}. Rather than evaluating agentic systems solely based on performance outcomes, organisations can use the autonomy score to ensure that decision authority remains aligned with organisational policies, risk tolerance, and regulatory requirements. For example, higher autonomy configurations may be acceptable in upstream planning environments characterised by high uncertainty, whereas downstream execution contexts may require stricter human oversight and lower tolerated autonomy levels. 

The results also provide insights into the interaction between autonomy and human involvement. Consultation rates are generally higher in upstream tiers, indicating increased reliance on human input under conditions of uncertainty. This suggests that human–AI interaction patterns are not uniform across the supply chain but instead vary with the informational characteristics of the decision environment. Consequently, governance mechanisms should account not only for the level of autonomy but also for where and when human intervention is most valuable. Designing effective human-in-the-loop systems therefore requires moving beyond static autonomy levels toward context-aware interaction policies \citep{cheng_toward_2026}.

From a managerial perspective, these findings have direct implications for the deployment of agentic AI in real-world supply chains. First, organisations should avoid assuming that increasing autonomy will automatically improve system performance. Instead, autonomy should be actively managed as part of a broader governance strategy. Second, deployment strategies should be context- and tier-specific, with results encouraging, counterintuitively, higher autonomy in upstream planning and coordination activities and more controlled autonomy in downstream, demand-facing operations. Third, continuous monitoring of autonomy through metrics such as the proposed autonomy score can support early detection of unintended behavioural shifts, enabling organisations to intervene before performance deteriorates. The application of the framework during initial development and the calibration of a desired autonomy range further enables the design of decision boundaries and policies to balance performance with resilience, flexibility and accountability.

Overall, the findings suggest that the primary value of the AAAA framework lies not in predicting performance, but in enabling controlled and transparent adoption of agentic AI systems. By providing a standardised and operational measure of autonomy, the framework supports the design of governance mechanisms that align agent behaviour with organisational objectives, thereby reducing the risks associated with increasingly autonomous decision-making in supply chains.

\section{Conclusions, Limitations and Future Work}
\label{sec:Conclusion}

Digital technologies have long shown the potential to enhance supply chain flexibility and resilience under increasing complexity. Recent advances in agentic AI further expand this potential by enabling systems that can operate autonomously, adapt to dynamic environments, and pursue long-horizon objectives. However, the governance required to support the effective and responsible adoption of such systems remains underdeveloped within the supply chain management literature.

This paper addresses this gap by proposing the Agentic AI Autonomy Assessment (AAAA) framework, which operationalises the degree of autonomy through observable task-level interactions. By distinguishing between delegation, consultation, and collaboration, the framework enables the continuous and objective measurement of autonomy throughout the lifecycle of an agentic system. In doing so, it shifts the conceptualisation of autonomy from an abstract system characteristic to a measurable and monitorable organisational variable.

The results from our experiments indicate that autonomy does not result in a uniform effect on performance outcomes. Instead, a position-dependent relationship is observed. These findings suggest that autonomy should not be interpreted as a universally desirable property, but rather as a context-dependent design parameter that interacts with system structure and informational dynamics. More broadly, the results reinforce the notion that autonomy is orthogonal to system capability, capturing how decisions are made rather than how well they are made \citep{bradshaw_dimensions_2004}. Consequently, increasing autonomy should not be expected to systematically improve operational performance, as decision outcomes remain dependent on numerous additional factors, including reasoning quality, domain knowledge, coordination strategies, environmental uncertainty, and the capabilities of the underlying AI model \citep{bradshaw_dimensions_2004, onnasch_human_2014}.

The study contributes to the literature on agentic AI and supply chain management in three ways. First, it introduces a novel, implementation-independent framework for measuring autonomy based on observable task interactions. Second, it provides an initial investigation into how autonomy interacts with supply chain dynamics, highlighting the importance of structural context. Third, it positions autonomy as a key dimension of governance, enabling organisations to define, monitor, and regulate the distribution of decision authority in agentic systems.

Several limitations should be considered when interpreting the findings. The empirical analysis is based on a simulated environment, which, while well-established, may not fully capture the complexity and variability of real-world supply chains. In addition, the experimental design relies on prompt-based manipulation of agent behaviour. Although this is similar to current implemented agentic systems, it may not fully control autonomy levels due to the inherent stochastic characteristics of large language models. This is reflected in the relatively narrow range of observed autonomy scores, suggesting limited sensitivity of the current experimental setup. Furthermore, the study focuses on delegation and consultation as primary interaction mechanisms, while the role of collaboration remains to be explored in greater depth.

These limitations open several avenues for future research. Further studies should validate the proposed framework in real-world or industry-grade agentic systems, as well as explore alternative experimental designs that provide stronger control over autonomy configurations. Expanding the framework to fully incorporate the dimension of collaboration and multi-agent coordination represents another important direction. Additionally, future work should investigate how autonomy interacts with other dimensions of system performance, such as robustness, adaptability, and decision quality, as well as how autonomy configurations across different supply chain tiers jointly influence overall system behaviour.

Overall, this study highlights that the primary value of autonomy assessment lies not in maximising performance, but in enabling its controlled and transparent deployment and enabling systemic monitoring over time. By providing a structured and continuous measure of autonomy, the AAAA framework offers a foundation for the development of governance mechanisms that can support the safe and effective integration of agentic AI into modern, digitalised supply chains.

%\subfile{0_Details}

\newpage

% \begin{thebibliography}{}
% \end{thebibliography}
\bibliography{mybib}

%\newpage

%\section{Appendices}
%\noindent\textbf{Appendix A. Include Full Agent Prompt?}\medskip

\end{document}